\documentclass[aps,prd,preprint,tightenlines,groupedaddress,showpacs]{revtex4}
\usepackage{epsf,epsfig,graphicx}
\begin{document}

\title{Affleck-Dine Baryogenesis, Split Supersymmetry, and Inflation}

\author{Yeo-Yie Charng$^1$}
\author{Da-Shin Lee$^2$}
\author{Chung Ngoc Leung$^3$}
\author{Kin-Wang Ng$^1$}
\affiliation{$^1$Institute of
Physics, Academia Sinica, Taipei,
Taiwan 115, R.O.C.\\
$^2$Department of Physics, National Dong Hwa University,
Hua-Lien, Taiwan 974, R.O.C.\\
$^3$Department of Physics and Astronomy, University of Delaware,
Newark, Delaware 19716, U.S.A.}

\vspace*{0.6 cm}
\date{\today}
\vspace*{1.2 cm}

\begin{abstract}
It is shown that, in the context of split supersymmetry, a simple
model with a single complex scalar field can produce chaotic
inflation and generate the observed amount of baryon asymmetry via
the Affleck-Dine mechanism. While the inflaton quantum
fluctuations give rise to curvature perturbation, we show that
quantum fluctuations of the phase of the scalar field can produce
baryonic isocurvature perturbation. Combining with constraints
from WMAP data, all parameters in the model can be determined to
within a narrow range.
\end{abstract}

\maketitle

\section{Introduction}

It is evident that no concentration of antimatter exists within
the Solar system and the Milky Way. The absence of annihilation
radiation from the Virgo cluster indicates that antimatter can
hardly be found within a 20 Mpc scale. A study of the contribution
of annihilation radiation near the matter-antimatter boundaries to
the cosmic diffuse gamma-ray background virtually excludes domains
of antimatter in the visible Universe~\cite{cohen}.

Direct observations of luminous matter show that baryons
constitute about 5 percents of the total mass of the Universe.
This gives a value of order $10^{-10}$ for $n_{\rm B}/s$, the
ratio of the baryon number density to the entropy density. Precise
measurements of the abundance of primordial light elements
predicted in big-bang nucleosynthesis combined with the cosmic
microwave background observations restrict this ratio in the
range~\cite{fields} $n_{\rm B}/s\simeq (4.7-6.5)\times 10^{-10}$.

Inflation~\cite{olive} is so far the most accepted paradigm for
understanding cosmological observations such as the flatness and
homogeneity of the observed Universe. More importantly, quantum
fluctuations of the inflaton can seed metric perturbations that
can subsequently grow to form cosmic structures. Despite this
success, the nature of the inflaton remains unknown.
Phenomenologically, the observed approximately scale-invariant
density power spectrum requires an inflaton that slowly rolls down
a nearly flat potential. The determination of the relevant model
and its potential would be a challenge to future observations.

When inflation ends, any pre-existing baryon asymmetry is washed
out. Many scenarios for baryon production after reheating or
during preheating have been proposed to explain the observed small
baryon asymmetry, such as baryogenesis in grand unified theories
(GUT), electroweak baryogenesis, leptogenesis, and Affleck-Dine
(AD) baryogensis~\cite{enqvist}. Affleck and Dine~\cite{afdine}
proposed a mechanism of baryogensis in supersymmetric (SUSY)
models in which scalar quark and lepton fields obtain large vacuum
expectation values along flat directions of the scalar potential.
These coherent scalars or the condensate start to oscillate when
SUSY-breaking effects start to become important, and a net baryon
number is developed and stored in the oscillating fields via
baryon-number violating dim-4 scalar couplings provided that $C$
and $CP$ symmetries are also violated. Subsequently, the scalar
quark and lepton fields decay and produce a baryon asymmetry.
However, in a certain AD flat direction the condensate is not the
state of lowest energy but fragments to form metastable or stable
Q-balls~\cite{enqvist}. Here we estimate the baryon asymmetry
assuming that the AD condensate does not lead to this type of
Q-ball formation.

In fact, the AD mechanism is too efficient and the resulting
baryon asymmetry is usually too large. Several dilution processes
have been considered to reduce the large AD baryon asymmetry to
the observed value.  They involve either introducing additional
entropy releases after baryogenesis (e.g., by the decays of the
inflaton~\cite{Linde}, the dilaton~\cite{dolgov}, or certain
massive scalar fields~\cite{kkk}) or reducing the baryon
production by invoking non-renormalizable terms~\cite{ng}. We
found in our recent work~\cite{charng} that non-equilibrium
effects, which had largely been ignored in earlier studies, could
play an important role in a certain parameter space and generate
the observed amount of baryon asymmetry, without any additional
dilution mechanism.

Recently, the model of split SUSY was proposed to avoid many
problems in SUSY standard model~\cite{split}. In the split SUSY
framework, the SUSY breaking scalar quark and lepton masses can be
as high as the GUT scale, while all the gauginos and Higgs bosons
are kept as light as TeV to facilitate both the converging of the
gauge couplings at the GUT scale and the lightest SUSY particle as
a viable dark matter candidate. Although an independent
fine-tuning of scale parameters is required to obtain an
acceptable ratio of vacuum expectation values
$\tan\beta$~\cite{drees}, the split SUSY scenario remains an
interesting possibility. This split mass spectrum has led to a new
consideration of the AD mechanism which showed that the smallness
of the baryon asymmetry is directly related to such a
spectrum~\cite{splitad}. In this paper, we propose a new scenario
in which the AD flat direction not only produces the baryon
asymmetry, but also plays the role of driving inflation.

\section{Affleck-Dine Mechanism}

Affleck and Dine~\cite{afdine} have shown that, in a SUSY SU(5)
grand unified model, there is a flat direction in the low-energy
effective potential for the following set of vacuum states of the
scalar up ($\tilde{u}$), strange ($\tilde{s}$), and bottom
($\tilde{b}$) quark fields as well as the scalar muon field
($\tilde{\mu}$):
\begin{eqnarray}
 &&\left<\tilde{u}_3^c\right>=A,\quad\quad
 \left<\tilde{u}_1\right>=B;\quad\quad
 \left<\tilde{s}_2^c\right>=A, \nonumber \\
 &&\left<\tilde{\mu}\right>=B;\quad\quad
 \left<\tilde{b}_1^c\right>=e^{i\xi}\sqrt{|A|^2+|B|^2},
\label{vev}
\end{eqnarray}
with all other fields having vanishing vacuum expectation values.
Here the superscript $c$ denotes charge conjugation, the
subscripts denote color indices, $A$ and $B$ are arbitrary complex
numbers, and $\xi$ is real.

To illustrate how the AD mechanism works, Affleck and
Dine~\cite{afdine} considered a toy model with a single complex
scalar field $\Phi$ described by the action
\begin{eqnarray}
 &&{\cal S} = \int d^4 x \, \sqrt{-g} \, \left[ \, g_{\mu \nu}
 \left(\partial^{\mu} \Phi^{\dagger} \right) \left(\partial^{\nu}
 \Phi \right) - V(\Phi)\right] \, , \nonumber \\
 &&V(\Phi)= m^2 \Phi^\dagger \Phi +
 i \lambda \left(\Phi^4 - \Phi^{\dagger\,4} \right),
\label{model}
\end{eqnarray}
which contains the $CP$ and baryon-number violating coupling,
$\lambda$. For example~\cite{afdine}, $\Phi$ represents the flat
direction associated with a combination of ${\bar u}{\bar d}{\bar
d}$ and $QL{\bar d}$, as shown in Eq.~(\ref{vev}), and the
$\lambda$-term is the $B-L=0$ dim-4 operator $Q{\bar u}^*L{\bar
d}^*$ arising from a one-loop box diagram. Another example is the
flat direction $QL{\bar d}$, which can be lifted by the dim-4
operator $QQQL$ that generates a nonvanishing A-term. The
background geometry is governed by the spatially flat
Robertson-Walker metric,
\begin{equation}
 d^2 s = d^2 t - a^2 (t) d^2 {\bf x} \,,
\end{equation}
where $a(t)$ is a scale factor. For small $\Phi$ and/or $\lambda$,
the theory has an approximately conserved current,
\begin{equation}
j_\mu = i \left( \Phi^\dagger \partial_\mu \Phi - (\partial_\mu
 \Phi^\dagger) \Phi \right),
\end{equation}
due to the approximate global $U(1)$ symmetry: $\Phi \rightarrow
e^{i \alpha} \Phi$. The corresponding charge will be referred to
as the baryon number:
\begin{equation}
j_0=n_B.
\end{equation}
Classically, in an expanding universe, the mean field ${\bar\Phi}$
obeys the equation of motion
\begin{equation}
 \ddot{\bar\Phi}+3H\dot{\bar\Phi}+m^2{\bar\Phi} =
 4i\lambda{\bar\Phi}^{\dagger\,3},
\label{eom}
\end{equation}
where $\dot{\bar\Phi} ={d{\bar\Phi}}/{dt}$, and $H={\dot a}/a$ is
the Hubble parameter. The classical particle number density is
given by
\begin{equation}
n_\Phi=m {\bar\Phi}^\dagger {\bar\Phi}. \label{nphi}
\end{equation}
The equation of motion~(\ref{eom}) implies that
\begin{equation}
 \dot n_B+3H n_B = -2{\rm Im}\left[{\bar\Phi}\frac{\partial
 V({\bar\Phi})}{\partial {\bar\Phi}}\right].
\label{nBeq}
\end{equation}
Here $n_B$ is the classical density defined in terms of the mean
field ${\bar\Phi}$.  At the time $t=t_0$ when $H(t)\simeq 2m/3$
and ${\bar\Phi}={\bar\Phi}_0$, ${\bar\Phi}$ starts to oscillate.
Then, expressing $\Phi$ in terms of the polar form,
\begin{equation}
\Phi=\frac{1}{\sqrt{2}} \phi e^{i\theta},
\end{equation}
the baryon number density can be approximated by
\begin{equation}
n_{B_0} \simeq -\frac{1}{m}{\rm Im}\left[{\bar\Phi}\frac{\partial
V({\bar\Phi})}{\partial{\bar\Phi}}\right]_{{\bar\Phi}={\bar\Phi}_0}
\simeq
-\frac{\lambda{{\bar\phi}_0}^4}{m}\cos\left(4{\bar\theta}_0\right).
\label{nB0final}
\end{equation}
Hence, the baryon number per particle is given by
\begin{equation}
\frac{n_B}{n_\Phi} \simeq \frac{n_{B_0}}{n_{\Phi_0}} \simeq
 -2\frac{\lambda{{\bar\phi}_0}^2}{m^2}
\cos\left(4{\bar\theta}_0\right). \label{rAD0}
\end{equation}
In the original AD mechanism~\cite{afdine}, $m^2 = M_S^2$ and
$\lambda = \gamma M_S^2/{M_G^2}$, where $M_S$ is the effective
SUSY breaking scale, $\gamma$ is a real parameter characterizing
$CP$ violation, and $M_G$ is the GUT scale. Hence, assuming
$\cos(4{\bar\theta}_0)\simeq O(1)$, $n_B/n_\Phi \simeq -\gamma
{{\bar\phi}_0}^2/M_G^2$, which can easily provide a large initial
$n_{\rm B}/s$ and thus dilution processes have to be introduced to
reduce it to the observed value. For example, taking
$\gamma=10^{-3}$, $M_S=10^{-16}M_P$, $M_G=10^{-2}M_P$, and
${{\bar\phi}_0}^2 = 10^{-3} M_P^2$, where $M_P$ is the Planck
mass, we find $\lambda=10^{-31}$ and $n_B/n_\Phi \simeq -10^{-2}$.
Note that the decay width of the condensate can be estimated
as~\cite{afdine}:
\begin{equation}
\Gamma_\Phi\sim \left(\frac{\alpha_s}{\pi}\right)^2
\frac{m^3}{|{\bar\Phi}|^2}, \label{decay}
\end{equation}
which is typically much smaller than the frequency of the
oscillating scalar fields, i.e., $\Gamma_\Phi\ll m$. Therefore, a
net baryon number is developed and gets saturated in the
oscillating scalar quark and lepton fields before they decay and
produce a baryon asymmetry.

In the following, we will discuss the AD mechanism in the context
of split SUSY and treat the potential~(\ref{model}) as the
inflaton potential. It turns out that this will limit the values
of the parameters $m$ and $\lambda$ to within a narrow range, thus
specifying a simple model that can both induce inflation
and produce the right amount of baryon asymmetry.

\section{Flat Direction and Inflation}

Let us introduce the real scalar fields $\sigma$ and $\chi$:
\begin{equation}
\frac{1}{\sqrt{2}}\left(\sigma+i\chi\right)=e^{-i{\bar\theta}_0}\Phi.
\end{equation}
In terms of these fields, the potential~(\ref{model}) becomes
\begin{equation}
V(\sigma,\chi)={1\over2}m^2\sigma^2+{1\over2}m^2\chi^2+\lambda\;{\rm
terms}. \label{chao}
\end{equation}
As long as $\lambda$ is sufficiently small, the classical
trajectory is well approximated by $\sigma \simeq \phi$ and $\chi
\simeq 0$.  Let us assume the AD $\sigma$-direction corresponds to
the inflaton. Then the potential~(\ref{chao}) is a typical
potential for chaotic inflation.

Since we identify the AD flat direction as the inflaton that has
an initial value larger than the Planck mass in the chaotic
inflation, the supergravity effects should be taken into account.
In the framework of supergravity, the effective potential is
modified as
\begin{equation}
V=e^{\frac{8\pi}{M_P^2}K}
\left[\left(\frac{\partial^2K}{\partial\Psi\partial\Psi^*}\right)^{-1}
D_\Psi W D_{\Psi^*} W^* - 3\frac{8\pi}{M_P^2} \vert W
\vert^2\right], \label{sugraV}
\end{equation}
with
\begin{equation}
D_\Psi W = \frac{\partial W}{\partial\Psi} + \frac{8\pi}{M_P^2}
\frac{\partial K}{\partial\Psi} W.
\end{equation}
Here $K$ is the K\"ahler potential, $W$ is the superpotential, and
$\Psi$ represents all relevant scalar fields in the model. In
minimal supergravity, the K\"ahler potential is given by
$K=\Psi\Psi^*$ (hereafter, for simplicity, we use the same
notations for superfields), so the exponential factor in
Eq.~(\ref{sugraV}) indeed prevents any fields from having values
larger than $M_P$ and the chaotic inflation would not occur.
However, there have been attempts to realize chaotic inflation in
the context of supergravity by using specific forms of the
K\"ahler potential~\cite{sugra,kawasaki}. Here we will adopt the
scenario for a natural chaotic inflation proposed in
Ref.~\cite{kawasaki}, in which the form of the K\"ahler potential
is determined by a shift symmetry of the inflaton field.

In a standard inflation model, the inflaton is in general a gauge
singlet, so it is not harmful to impose the shift symmetry on it.
In the present consideration, the shift symmetry imposed on the
flat direction, which are scalar quark and lepton fields, is
apparently incompatible with the standard-model gauge groups. This
can be seen in the terms (see below) that we will introduce in the
K\"ahler potential~(\ref{Kpot}) and the
superpotential~(\ref{Spot}). However, in the AD mechanism, the
scalar quark and lepton fields get vacuum expectation values which
break the standard-model gauge symmetries. As such, during
inflation there may exist effective operators, for examples,
${\bar u}^*{\langle Q\rangle}{\langle L\rangle}{\langle{\bar
d}^*\rangle}$ and ${\bar u}^*L{\langle Q\rangle}{\langle{\bar
d}^*\rangle}$, which do not respect the standard-model gauge
symmetries. This particular feature of the AD mechanism indeed
opens a possibility for imposing a shift symmetry on the AD flat
direction even though the flat direction is not a gauge singlet.
For instance, assuming that all the vacuum expectation values are
real numbers, the Lagrangian may contain terms like ${\bar
u}+{\bar u}^*$ which carries the shift symmetry. Interestingly,
these operators are not harmful at all since they will vanish once
the scalar quark and lepton fields settle to the ground state,
i.e., their vacuum expectation values approaches zero. In light of
this, it is not impossible to make use of the shift symmetry to
construct a slow-roll condition for the AD flat direction in the
context of supergravity. Below we will present a schematic way for
the construction instead of deriving it from the full theory. The
full derivation is very interesting and it certainly warrants a
further detailed investigation.

With respect to the flat direction~(\ref{vev}), we assume the
following K\"ahler potential:
\begin{equation}
K(A,B)={1\over2}(A+A^*)^2 + BB^*, \label{Kpot}
\end{equation}
which is invariant under the shift of $A$: $A \rightarrow
A+icM_P$, where $c$ is a real parameter. As a consequence, the
exponential factor in Eq.~(\ref{sugraV}) no loner prevents the
imaginary part of $A$ from having a larger value than $M_P$, which
we identify with the inflaton field $\sigma$ in the
potential~(\ref{chao}). As long as the superpotential is given by
a quadratic mass term:
\begin{equation}
W=mAB, \label{Spot}
\end{equation}
we find that the effective potential is given by~\cite{kawasaki}
\begin{eqnarray}
V(A,B)&=& m^2 e^{\frac{8\pi}{M_P^2}K} \left\{ \vert
A\vert^2\left[1+\left(\frac{8\pi}{M_P^2}\right)^2\vert
B\vert^4\right] \right.
\nonumber \\
&& \left. + \vert B\vert^2 \left[1-\frac{8\pi}{M_P^2}\vert
A\vert^2 +\frac{8\pi}{M_P^2}(A+A^*)^2
\left(1+\frac{8\pi}{M_P^2}\vert A\vert^2\right)\right]\right\}.
\label{effV}
\end{eqnarray}
Now, let us express $A$ and $\vert B\vert$ in terms of two real
rectilinear components and a real radial component respectively as
\begin{equation}
A=\frac{1}{\sqrt2}\left(\rho+i\sigma\right),\quad \vert
B\vert=\frac{1}{\sqrt2}\chi,
\end{equation}
then the effective potential~(\ref{effV}) becomes
\begin{eqnarray}
V(\rho,\sigma,\chi)&=& m^2
e^{\frac{8\pi}{M_P^2}\left(\rho^2+\chi^2/2\right)} \left\{
{1\over2}\left(\rho^2+\sigma^2\right)\left[1+\left(\frac{8\pi}{M_P^2}\right)^2
{1\over4}\chi^4\right] \right.
\nonumber \\
&& \left. + {1\over2}\chi^2
\left[1-\frac{8\pi}{M_P^2}{1\over2}\left(\rho^2+\sigma^2\right)
+\frac{8\pi}{M_P^2}2\rho^2 \left(1+\frac{8\pi}{M_P^2}
{1\over2}\left(\rho^2+\sigma^2\right)\right)\right]\right\}.
\end{eqnarray}
Since the exponential factor contains $\rho$ and $\chi$, they are
refrained from having values larger than $M_P$. On the contrary,
$\sigma$ can take a value much larger than $M_P$. For
$\vert\rho\vert,\vert\chi\vert\ll M_P$, we can expand the
potential. Keeping terms up to second order in the
supergravity corrections, we obtain
\begin{eqnarray}
V(\rho,\sigma,\chi)&\simeq& {1\over2}m^2\sigma^2 +
{1\over2}m^2\chi^2\left[1+\frac{8\pi}{M_P^2}{1\over2}\chi^2+
\left(\frac{8\pi}{M_P^2}\right)^2{1\over8}\chi^4\right]
+\left(\frac{8\pi}{M_P^2}\right)^2{m^2\over16}\sigma^2\chi^4
\nonumber\\
&&+{1\over2}m^2\rho^2\left[1+\frac{8\pi}{M_P^2}\left(\sigma^2+3\chi^2+\rho^2\right)\right.
\nonumber \\
&&\left.+\left(\frac{8\pi}{M_P^2}\right)^2\left(\sigma^2\chi^2+{1\over2}\sigma^2\rho^2+{13\over8}\chi^4
+{7\over2}\chi^2\rho^2 +{1\over2}\rho^4\right)\right].
\label{effV2}
\end{eqnarray}
In chaotic inflation, the initial value of the inflaton $\sigma$
is larger than $M_P$ (see below). For such a large value of
$\sigma$, the effective mass of $\rho$ becomes much larger than
$m$, so its quantum fluctuations are suppressed~\cite{kkk2}.
Therefore, it is legitimate to assume that $\rho$ behaves like a
classical field sitting at the origin. With $\rho=0$ and
$\vert\chi\vert\ll M_P$, Eq.~(\ref{effV2}) effectively becomes the
potential for chaotic inflation as given in Eq.~(\ref{chao}). The
$\chi^4, \chi^6, \sigma^2\chi^4$ terms in Eq.~(\ref{effV2}) only
slightly modify the number density and do not affect the baryon
number (see Eqs.~(\ref{nphi}) and~(\ref{nBeq})).

Another way of taming the supergravity effects requires the
introduction of a modulus field $Z$ and a specific choice of the
K\"ahler potential~\cite{murayama}:
\begin{equation}
K=\frac{M_P^2}{8\pi}\left({3\over8}\ln y+y^2\right),\quad
y=\frac{8\pi}{M_P^2}\left(Z+Z^*+\Phi^*\Phi\right). \label{kzphi}
\end{equation}
so that the scalar potential for a given superpotential $W(\Phi)$
reads~\cite{murayama}
\begin{equation}
V=y^{3\over8} e^{y^2} \left[\frac{8y}{16y^2+3} \left\vert
\frac{\partial W}{\partial\Phi} \right\vert^2 +
\frac{8\pi}{M_P^2}\frac{(16y^2-9)^2}{8(16y^2-3)} \vert W\vert^2
\right].
\end{equation}
For $W=m\Phi^2/2$, it was shown~\cite{murayama} that $y$ settled
quickly to the value $y=3/4$ during inflation and the potential became
\begin{equation}
V=\left({3\over4}\right)^{3\over8} e^{9\over16} {1\over2}
\left\vert \frac{\partial W}{\partial\Phi} \right\vert^2\simeq m^2
\vert \Phi \vert^2.
\end{equation}
Although the slow-roll condition is maintained, the choice of the
K\"ahler potential~(\ref{kzphi}) is not motivated by any symmetry
argument. A possibility is to impose a Heisenberg symmetry on the
K\"ahler potential. In Ref.~\cite{gaillard}, a combination of $Z$
and $\Phi$ which is invariant under the Heisenberg symmetry is
constructed as
\begin{equation}
I=Z+Z^*-\Phi^*\Phi,
\end{equation}
and the K\"ahler potential is assumed to be a function of $I$
only:
\begin{equation}
K=\frac{M_P^2}{8\pi}f(\bar I),\quad {\bar I}=\frac{8\pi}{M_P^2} I.
\end{equation}
It can then be shown that $I$ and $\Phi$ are independent degrees
of freedom and the scalar potential for a given superpotential
$W(\Phi)$ is given by
\begin{equation}
V=e^{f(\bar I)}
\left[\frac{8\pi}{M_P^2}\left(\frac{f'^2}{f''}-3\right)\vert
W\vert^2 - \frac{1}{f'} \left\vert \frac{\partial W}{\partial\Phi}
\right\vert^2 \right], \label{viphi}
\end{equation}
where the prime denotes differentiation with respect to $\bar I$.
For the no-scale model with $f=-3\ln{\bar I}$, $f'^2=3f''$ and the
first term in Eq.~(\ref{viphi}) vanishes. The potential then takes
the form
\begin{equation}
V={1\over3}e^{{2\over3}f} \left\vert \frac{\partial
W}{\partial\Phi} \right\vert^2,
\end{equation}
and it is usually assumed that higher-order corrections can
stabilize the $\bar I$ field during inflation~\cite{gaillard}. As
long as the $\bar I$ field is constant during inflation and the
superpotential is $W=m\Phi^2/2$, chaotic inflation is
preserved.

In Eq.~(\ref{chao}), we will deal with a small $\lambda$ such that
the mass term dominates the classical motion: $V \simeq
m^2\sigma^2/2$. Define the slow-roll parameters in terms of the
mean field $\bar\sigma$:
\begin{equation}
\epsilon\equiv \frac{M_P^2}{16\pi}\left({1\over V}\frac{\partial
V}{\partial {\bar\sigma}}\right)^2,\quad \eta\equiv
\frac{M_P^2}{8\pi}\left({1\over V}\frac{\partial^2 V}{\partial
{\bar\sigma}^2}\right),
\label{srparameters}
\end{equation}
where $\partial V/\partial {\bar\sigma}$ stands for $\partial V/
\partial \sigma$ evaluated at $\sigma = \bar\sigma$. 
In the $\lambda = 0$ limit, 
\begin{equation}
\epsilon=\eta=\frac{M_P^2}{4\pi{\bar\sigma}^2}.
\label{srpara2}
\end{equation}
In addition, the equations of motion are given by
\begin{eqnarray}
&&H^2\equiv \left(\frac{\dot a}{a}\right)^2 =\frac{8\pi}{3M_P^2}
\left({1\over2}{\dot{\bar\sigma}}^2+V\right),
\label{hubble} \\
&& \frac{\ddot a}{a}=\frac{8\pi}{3M_P^2}
\left(-{\dot{\bar\sigma}}^2+V\right),
\label{acc} \\
&&\ddot{\bar\sigma}+3H\dot{\bar\sigma}+ \frac{\partial
V}{\partial {\bar\sigma}}= 0.
\label{sigmaeom}
\end{eqnarray}
During the slow-roll stage, the number of e-folds from the end of
inflation is
\begin{equation}
N=\int_t^{t_e} H dt \simeq
\frac{2\pi}{M_P^2}\left({\bar\sigma}^2-{\bar\sigma}_e^2\right),
\quad {\bar\sigma}_e^2=\frac{M_P^2}{3\pi}, \label{sigmaend}
\end{equation}
where ${\bar\sigma}_e$ is the field value as it starts to
oscillate. Hence, the slow-roll parameter $\epsilon$ can be 
approximated by
\begin{equation}
\epsilon\simeq
\frac{1}{2N}\left(1+\frac{2\pi{\bar\sigma}_e^2}{NM_P^2}\right)^{-1}.
\label{epsilon}
\end{equation}
We shall find it useful in the following to express the slow-roll
dynamics in terms of the slow-roll parameters $\epsilon$ and
$\eta$.  In the slow-roll approximation,
\begin{equation}
\ddot{\bar\sigma}={1\over 3}\frac{\partial V}
{\partial {\bar\sigma}}(\eta-\epsilon)
\end{equation}
is negligibly small.  Eqs.~(\ref{hubble}), (\ref{acc}), and
(\ref{sigmaeom}) then imply
\begin{eqnarray}
H^2&\simeq&\frac{8\pi V}{3M_P^2}
\left(1+\frac{\epsilon}{3}\right) \simeq \frac{m^2}{3\epsilon}, 
\label{hubblesr}\\
\frac{\ddot a}{a}&\simeq&H^2 \left(1-\epsilon\right).
\end{eqnarray}
It is now straightforward to show that
\begin{equation}
\frac{\dot H}{H^2} \simeq -\epsilon.
\end{equation}
Defining the conformal time $d\tau=dt/a$, $z=a{\dot{\bar\sigma}}/H$,
and $z'=dz/d\tau$, we obtain from the above equations that
\begin{eqnarray}
a(\tau)&=&-\frac{1}{H\tau(1-\epsilon)}, \label{atau} \\
\frac{a''}{a}&=&\frac{1}{\tau^2}(2+3\epsilon), \label{apprime} \\
\frac{z''}{z}&=&\frac{1}{\tau^2}(2+9\epsilon-3\eta). \label{zpprime}
\end{eqnarray}

\section{Density Perturbation}

Now we turn to consider the density perturbation generated during
inflation. There are two kinds of density perturbation. Quantum
fluctuations of the $\sigma$ field induce the adiabatic density
perturbation, whereas those of $\chi$ do not affect the energy
density and give rise to isocurvature perturbation. Here we will
follow the discussions in Ref.~\cite{byrnes}.

In the spatially-flat gauge, the Fourier mode of $\sigma$
fluctuations, $\delta\sigma$, obeys
\begin{equation}
\ddot{\delta\sigma}+3H\dot{\delta\sigma}+\left[\frac{k^2}{a^2}
+\frac{\partial^2 V}{\partial{\bar\sigma}^2}
-\frac{8\pi}{M_P^2 a^3} \frac{d}{dt}
\left(\frac{a^3\dot{\bar\sigma}^2}{H}\right)\right]{\delta\sigma}=0.
\end{equation}
Expressing the above equation in terms of the slow-roll
parameters
and using Eq.~(\ref{zpprime}), we obtain
\begin{equation}
u''+\left(k^2-\frac{z''}{z}\right)=0,
\end{equation}
where $u=a\delta\sigma$, $u'=du/d\tau$, and $z=a{\dot{\bar\sigma}}/H$.
When $k^2\gg z''/z\simeq 2a^2H^2$, $u$ has a plane-wave solution,
\begin{equation}
u\simeq\frac{e^{ik\tau}}{\sqrt{2k}}.
\end{equation}
When $k^2\ll z''/z$, we can write
\begin{equation}
u\simeq A(k) z.
\end{equation}
The spectral function $A(k)$ is determined by
\begin{equation}
A(k)\simeq \left[\frac{u}{z}\right]_{k=aH}=
\left[\frac{u}{a}\frac{H}{\dot{\bar\sigma}}\right]_{k=aH}
=\left[\frac{H}{a\dot{\bar\sigma}}\frac{e^{ik\tau}}{\sqrt{2k}}\right]_{k=aH},
\end{equation}
where the quantities inside the square brackets are evaluated at the
time of horizon-crossing. Hence, the power spectrum of $\sigma$ fluctuations is given by
\begin{equation}
P_\sigma=\frac{4\pi k^3}{(2\pi)^3}\vert\delta\sigma\vert^2
\simeq \frac{4\pi k^3}{(2\pi)^3} \left(\frac{\dot{\bar\sigma}}{H}\right)^2 \vert A(k)\vert^2
=\left(\frac{\dot{\bar\sigma}}{H}\right)^2
 \left[\frac{H}{\dot{\bar\sigma}} \frac{H}{2\pi} \right]^2_{k=aH}.
\end{equation}
Using Eqs.~(\ref{srpara2}), (\ref{hubble}), and (\ref{sigmaeom}),
we find that the adiabatic density perturbation is described by
\begin{equation}
P_\zeta = \left(\frac{H}{\dot{\bar\sigma}}\right)^2 P_\sigma
\simeq \frac{16\pi}{3} \left[\frac{m{\bar\sigma}^2}{M_P^3}\right]^2_{k=aH}
\simeq \frac{m^2}{3\pi M_P^2} \left[\frac{1}{\epsilon^2}\right]_{k=aH}. \label{Pzeta}
\end{equation}
To have enough inflation, it is required that~\cite{liddle}
$N\simeq 60$. From Eq.~(\ref{epsilon}), $\epsilon\simeq
0.008$. Hence, the WMAP measurement of the matter power
spectrum~\cite{wmap}, $P_\zeta\simeq 2\times 10^{-9}$, implies
that $m\simeq 1.1\times 10^{-6} M_P$.  The tilt of the adiabatic
density spectral index is then given by
\begin{equation}
\Delta n_\zeta\equiv \frac{d\ln P_\zeta}{d\ln k}\simeq
-6\epsilon+2\eta=-4\epsilon\simeq -0.03,
\end{equation}
which is consistent with WMAP measurements~\cite{wmap}.

Since $\chi=0$ in the background solution, the isocurvature field
fluctuations $\delta\chi$ are gauge-independent and the Fourier
mode satisfies the massless Klein-Gordon equation in de Sitter
space,
\begin{equation}
v''+\left(k^2+a^2m^2-\frac{a''}{a}\right)v=
v''+\left(k^2-\frac{2}{\tau^2}\right)v=0,
\end{equation}
where $v=a\delta\chi$ and we have used Eqs.~(\ref{atau}), (\ref{apprime}) and (\ref{hubblesr}).
Taking the Bunch-Davis vacuum for the
solution $v$, we have
\begin{equation}
v\simeq\frac{e^{ik\tau}}{\sqrt{2k}}
\left(1+\frac{i}{k\tau}\right).
\end{equation}
Noting that $\delta\chi$ evolves exactly like ${\bar\sigma}$ in
Eq.~(\ref{sigmaeom}) for $k^2\ll a^2 m^2$, we can write
\begin{equation}
\delta\chi\simeq B(k) {\bar\sigma},
\end{equation}
where the spectral function $B(k)$ is determined by
\begin{equation}
B(k)\simeq \left[\frac{v}{a{\bar\sigma}}\right]_{k=aH}\simeq
\left[\frac{1}{a{\bar\sigma}}\frac{e^{ik\tau}}{\sqrt{2k}}
\left(\frac{i}{k\tau}\right)\right]_{k=aH}.
\end{equation}
Thus, the power spectrum of isocurvature fluctuations is found as
\begin{equation}
P_\chi=\frac{4\pi k^3}{(2\pi)^3}\vert\delta\chi\vert^2
\simeq \frac{4\pi k^3}{(2\pi)^3} {\bar\sigma}^2\vert B(k)\vert^2
\simeq {\bar\sigma}^2 \left[\frac{(1-\epsilon)H}{2\pi{\bar\sigma}}
\right]^2_{k=aH},
\end{equation}
where Eq.~(\ref{atau}) has been used.  In addition, Eq.~(\ref{hubblesr}) gives
\begin{equation}
H^2 \simeq \frac{4\pi}{3}
\frac{m^2{\bar\sigma}^2}{M_P^2}\left(1+\frac{\epsilon}{3}\right).
\end{equation}
The isocurvature power spectrum therefore becomes
\begin{equation}
P_\chi\simeq \frac{1}{3\pi} \frac{m^2{\bar\sigma}^2}{M_P^2}
\left(1-\frac{5\epsilon}{3}\right), \label{Pchi}
\end{equation}
and the isocurvature spectral tilt is
\begin{equation}
\Delta n_\chi\equiv \frac{d\ln P_\chi}{d\ln k}\simeq
-{10\over3}\epsilon^2\simeq -0.0002,
\end{equation}
which is nearly scale invariant.

\section{Baryogenesis and Baryonic Isocurature Perturbation}

After inflation, the inflaton or the AD condensate starts
oscillating and the oscillating field carries a baryon asymmetry
found in Eq.~(\ref{nB0final}), where ${\bar\phi}_0$ is now the
field value at the end of inflation, ${\bar\sigma}_e$, as given in
Eq.~(\ref{sigmaend}) (the time $t_e$ corresponds to the time of
baryogenesis, $t_0$). During this period, the Universe is
matter-dominated by the oscillating condensate and the expansion
rate is given by time-averaging Eq.~(\ref{hubble}) over field
oscillations:
\begin{equation}
H=\left[\frac{8\pi}{3M_P^2}
\left\langle{1\over2}{\dot{\bar\sigma}}^2
+{1\over2}m^2{\bar\sigma}^2\right\rangle\right]^{1\over2} =
\sqrt{\frac{8\pi}{3}}\frac{m}{M_P}
\langle{\bar\sigma}^2\rangle^{1\over2}. \label{hubble2}
\end{equation}
However, the baryon number per particle is constant, given by
Eq.~(\ref{rAD0}) as
\begin{equation}
\frac{n_B}{n_\Phi} \simeq -\frac{2}{3\pi} \frac{\lambda
M_P^2}{m^2} \cos\left(4{\bar\theta}_0\right). \label{rAD}
\end{equation}

Comparing the expansion rate~(\ref{hubble2}) with the decay
rate~(\ref{decay}), we find that the condensate decays into light
quarks and leptons when
\begin{equation}
\langle{\bar\sigma}^2\rangle^{1\over2}_d =
\left(\frac{3\alpha_s^4}{2\pi^5}\right)^{1\over6}
\left(m^2M_P\right)^{1\over3}. \label{sigmad}
\end{equation}
When the condensate decays, the decay particles are relativistic
and each carries an energy of the order of $m$. They will get
thermalized through the scattering process. At this time, the
elastic scattering cross section $\sigma_T=\alpha_G^2/m^2$ and the
thermalization rate is given by
\begin{equation}
\Gamma_T\sim {n_\Phi}_d \sigma_T
={1\over2}m\langle{\bar\sigma}^2\rangle_d \frac{\alpha_G^2}{m^2}
=\left(\frac{9\alpha_s^2\alpha_G^6}{256\pi^4}\right)^{1\over3}
\left(\frac{M_P}{m}\right)^{4\over3} H_d,
\end{equation}
where ${n_\Phi}_d$ and $H_d$ are respectively the approximate
number density of decay particles and the Hubble parameter at the
decay. For $\alpha_s^2\simeq\alpha_G^2\simeq 10^{-3}$ and $m\simeq
1.1\times 10^{-6} M_P$, we find that $\Gamma_T\simeq 627 H_d$,
implying that the decay products are thermalized instantly
relative to the expansion time. As such, the reheating temperature
of the thermalized radiation can be estimated as
\begin{equation}
\frac{\pi^2}{30} N(T_{re}) T_{re}^4 =
m^2\langle{\bar\sigma}^2\rangle_d,
\end{equation}
where $N(T_{re})$ counts the effective degrees of freedom. From
Eq.~(\ref{sigmad}), we find that
\begin{equation}
T_{re}=\left(\frac{40500\alpha_s^4}{\pi^{11}}\right)^{1\over12}
N^{-{1\over4}}(T_{re}) \left(\frac{M_P}{m}\right)^{1\over6} m.
\end{equation}
Assuming that $N(T_{re})\simeq 10^2$, then $T_{re}\simeq 0.834
m\simeq 1.1\times 10^{13} {\rm GeV}$.

\begin{figure}[t]
\begin{center}
\includegraphics[scale=0.45]{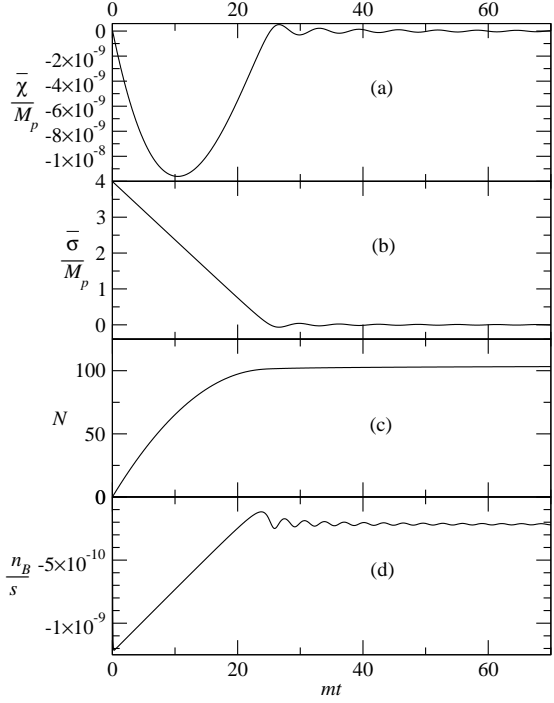}
\caption{Time evolution of (a) ${\bar\chi}(t)$, (b)
${\bar\sigma}(t)$, (c) the number of e-folds $N(t)$, and (d)
$n_B/s$. At $mt\simeq 25$, chaotic inflation ends and the baryon
asymmetry is generated.} \label{fig1}
\end{center}
\end{figure}

Although the reheating temperature $T_{re}$ is high, the
cosmological gravitino problem can be easily avoided. It is
because the gravitino mass, $m_{3/2}$, can be as large as the mass
scale $m$. For $m_{3/2}\ll m$, gravitinos may dominate the
Universe and produce extra entropy at the decay, thus diluting the
baryon asymmetry produced by the AD condensate. This light
gravitino case has been discussed in detail in
Ref.~\cite{splitad}. It was found that when $m_{3/2}< 10^9 {\rm
GeV}$, the extra entropy production dilutes the baryon asymmetry
by a dilution factor which is about $(m_{3/2}/10^9 {\rm
GeV})^{-5/2}$. In the following, we assume that $m_{3/2} > 10^9
{\rm GeV}$. In the case of $m_{3/2} < 10^9 {\rm GeV}$, one will
have to take the dilution factor into account. Then, the entropy
at reheating is given by
\begin{equation}
s=\frac{2\pi^2}{45} N(T_{re}) T_{re}^3
={4\over3}\frac{m^2\langle{\bar\sigma}^2\rangle_d}{T_{re}}
\end{equation}
and the baryon number density is given by the particle number
density of the decaying AD condensate times the ratio found in
Eq.~(\ref{rAD}):
\begin{equation}
n_B=-{1\over2}m\langle{\bar\sigma}^2\rangle_d \frac{2}{3\pi}
\frac{\lambda M_P^2}{m^2} \cos\left(4{\bar\theta}_0\right).
\end{equation}
Hence, the baryon to entropy ratio is
\begin{equation}
\frac{n_B}{s} \simeq -\frac{1}{4\pi} \frac{T_{re}}{m}\frac{\lambda
M_P^2}{m^2} \cos\left(4{\bar\theta}_0\right). \label{rADfinal}
\end{equation}
Furthermore, the baryonic isocurvature perturbation induced by
$\delta\chi$ during inflation can be derived as
\begin{equation}
\frac{\delta{n_{B}}_0}{{n_{B}}_0}\simeq -4
\tan\left(4{\bar\theta}_0\right)
\frac{\delta\chi_e}{{\bar\sigma}_e}
\end{equation}
and the corresponding power spectrum is
\begin{equation}
P_B=\frac{4\pi k^3}{(2\pi)^3}
\left(\frac{\delta{n_{B}}_0}{{n_{B}}_0}\right)^2 \simeq
\frac{16}{{\bar\sigma}_e^2} \tan^2\left(4{\bar\theta}_0\right)
P_{\chi_e}, \label{PB}
\end{equation}
where $P_{\chi_e}$ is given by Eq.~(\ref{Pchi}) with
${\bar\sigma}={\bar\sigma}_e$. Defining a density-weighted ratio
of the isocurvature and adiabatic perturbations, and using
Eqs.~(\ref{Pzeta}) and (\ref{PB}), we have
\begin{equation}
\kappa\equiv \frac{\Omega_B}{\Omega_m}\sqrt{\frac{P_B}{P_\zeta}}
\simeq 4\frac{\Omega_B}{\Omega_m} \tan\left(4{\bar\theta}_0\right)
\epsilon, \label{kiso}
\end{equation}
where $\Omega_B$ and $\Omega_m$ are, respectively, the present
baryon and matter densities relative to the critical density. Note
that $\kappa$ is suppressed by the slow-roll parameter $\epsilon$.
Combining Eqs.~(\ref{kiso}) and (\ref{rADfinal}), we obtain
\begin{equation}
\lambda\simeq 4\pi \frac{m}{T_{re}}\frac{m^2}{M_P^2} \frac{n_B}{s}
\sqrt{1+\left(\frac{\kappa}{4\epsilon}\frac{\Omega_m}{\Omega_B}
\right)^2}.
\end{equation}
>From above, $m\simeq 1.1\times 10^{-6} M_P$, $\epsilon\simeq
0.008$, and $T_{re}\simeq 0.834m$. Using the WMAP cosmological
parameters~\cite{spergel}, $\Omega_B=0.044$, $\Omega_m=0.268$, and
$n_B/s\simeq 10^{-10}$, the upper limit set on the amount of
isocurvature perturbation by WMAP data~\cite{hama}, $\kappa<0.4$
at $95\%$ confidence level, implies that
\begin{equation}
1.8\times 10^{-21}< \lambda < 1.6\times 10^{-20}.
\label{inequality}
\end{equation}
According to the result of Ref.~\cite{charng}, nonequilibrium
effects are negligible for such small $\lambda$ values.  We
therefore expect the mean-field approximation used in the above
analysis to be valid. In the case of $m_{3/2} < 10^9 {\rm GeV}$,
the baryon asymmetry~(\ref{rADfinal}) is diluted by a factor of
$(m_{3/2}/10^9 {\rm GeV})^{-5/2}$. Thus, $\lambda$ in
Eq.~(\ref{inequality}) is replaced by $\lambda (m_{3/2}/10^9 {\rm
GeV})^{5/2}$.

In Fig~\ref{fig1}, we show the results from a numerical
calculation of the chaotic inflation and AD baryogenesis, using
the model~(\ref{chao}) with $m\simeq 1.1\times 10^{-6} M_P$,
$\lambda\simeq 1.8\times 10^{-21}$, and the initial conditions,
$\bar\sigma=4M_P$ and
$\bar\chi=\dot{\bar\sigma}=\dot{\bar\chi}=0$. This model gives an
inflation of about $100$ e-folds as well as a baryon asymmetry in
good agreement with the approximation~(\ref{rADfinal}). Note that
near the beginning of inflation $n_B/s$ decreases rapidly from
zero to a minimum that can be estimated from Eq.~(\ref{nBeq})
(when $\dot n_B=0$) to be $-4\lambda{\bar\sigma}^2/(3mH)$, where
$\bar\sigma=4M_P$ and $H$ is given by Eq.~(\ref{hubble}) evaluated
at $t\simeq 0$.

\section{Conclusion}

In conclusion, we find that the heavy scalar fermions in split
supersymmetry can simultaneously play the role of driving
inflation and produce the baryon asymmetry of the Universe, as
long as the scalar mass $m\simeq 10^{13}$ GeV and the CP-violating
self-coupling $\lambda \simeq 10^{-20}$.

Upper limits on the scale of the lightest scalar quark mass coming
from cosmological constraints on a long-lived gluino have been
explored~\cite{gluino}.  For gluinos heavier than a few hundred
GeV the constraints from big bang nucleosynthesis set an upper
bound on the scalar quark mass of about $10^9$ GeV.  For lighter
gluinos, the constraints are from non-observation of diffuse gamma
rays and set an upper bound of about $10^{12}$ GeV. Although the
parameter $m$ in our model is not necessarily the lightest scalar
quark mass, it is interesting that the required value of $m$ is
close to this upper bound.  We are not aware of any constraint on
$\lambda$, whereas the expected theoretical value is given by the
size of the $A$-term.  In split SUSY, the gaugino masses are
suppressed by some mechanism, which also keeps a small $A\sim {\rm
TeV}$, as in the D-breaking scenario~\cite{split}.  As such,
$\lambda\sim \gamma A/M_P\sim  10^{-19}$, which is close to the
required value above. Interestingly, our scenario may provide
useful clues for building split SUSY models and have implications
for gluino searches at the LHC. In addition, the generated
baryonic isocurvature perturbation is nearly scale-invariant and
may saturate the present CMB observational upper limit.  Upcoming
Planck CMB mission will improve this limit and further test our
scenario.

We have also constructed a slow-rolling inflaton potential for the
Affleck-Dine field in the context of supergravity. It will be
interesting to include nonrenormalizable terms and to consider a
full model in supergravity as discussed, for example, in Dine
{\it et al.} in Ref.~\cite{ng}. However, since the scalar fermion
is the inflaton itself and $m\sim H$, we do not expect a drastic
modification, as supported by the results in Ref.~\cite{splitad}.

\begin{acknowledgments}
This work was supported in part by the NSC, ROC, under grants
NSC95-2112-M-001-052-MY3 and NSC95-2112-M-259-011-MY2, and by the
U.S. Department of Energy under grant DE-FG02-84ER40163. CNL would
like to thank the NCTS and the Academia Sinica for their
hospitality and support for collaboration visits.
\end{acknowledgments}


\begin{thebibliography}{99}
\bibitem{cohen}
A. G. Cohen, A. De Rujula, and S. L. Glashow,
Astrophys. J. {\bf 495}, 539 (1998).

\bibitem{fields}
B. D. Fields and S. Sarkar, in {\it The Review of Particle
Physics}, edited by C. Amsler {\it et~al.}, Phys. Lett. B {\bf
667}, 1 (2008).

\bibitem{olive}
For reviews see:  K. A. Olive, Phys. Rep. {\bf 190}, 307 (1990);
D. H. Lyth and A. Riotto, Phys. Rep. {\bf 314}, 1 (1999).

\bibitem{enqvist}
For a review, see K. Enqvist and A. Mazumdar, Phys. Rep. {\bf
380}, 99 (2003).  See also J. McDonald and O. Seto, J. Cosmol.
Astropart. Phys. 07 (2008) 015.

\bibitem{afdine}
I. Affleck and M. Dine, Nucl. Phys. B {\bf 249}, 361 (1985).

\bibitem{Linde}
A. D. Linde, Phys. Lett. B {\bf 160}, 243 (1985); J.~R. Ellis,
K.~Enqvist, D.~V. Nanopoulos, and K.~A. Olive, Phys. Lett. B {\bf
191}, 343 (1987).

\bibitem{dolgov}
A.~D. Dolgov, K. Kohri, O. Seto, and J. Yokoyama, Phys. Rev. D
{\bf 67}, 103515 (2003).

\bibitem{kkk}
K. Enqvist, K.-W. Ng, and K. A. Olive, Phys. Rev. D {\bf 37}, 2111 (1988).

\bibitem{ng}
K.-W. Ng, Nucl. Phys. B {\bf 321}, 528 (1989); M. Dine, L.
Randall, and S. Thomas,  Nucl. Phys. B {\bf 458}, 291 (1996);
B.~A. Campbell, M.~K. Gaillard, H. Murayama, and K.~A. Olive,
Nucl. Phys. B {\bf 538}, 351 (1999).

\bibitem{charng}
Y.-Y. Charng, D.-S. Lee, C. N. Leung, and K.-W. Ng, Phys. Rev. D
{\bf 72}, 123517 (2005).

\bibitem{split}
N. Arkani-Hamed and S. Dimopoulos, J. High Energy Phys. {\bf 06},
073 (2005); G. F. Giudice and A. Romanino, Nucl. Phys. B {\bf
699}, 65 (2004); N. Arkani-Hamed, S. Dimopoulos, G.~F. Giudice,
and A. Romanino, Nucl. Phys. B {\bf 709}, 3 (2005).

\bibitem{drees}
M. Drees, arXiv:hep-ph/0501106.

\bibitem{splitad}
S. Kasuya and F. Takahashi, Phys. Rev. D {\bf 71}, 121303 (2005).

\bibitem{sugra}
A. S. Goncharov and A. D. Linde, Phys. Lett. B {\bf 139}, 27
(1984); Classical Quantum Gravity {\bf 1}, L75 (1984); H.
Murayama, H. Suzuki, T. Yanagida, and J. Yokoyama, Phys. Rev. D
{\bf 50}, R2356 (1994).

\bibitem{kawasaki}
M. Kawasaki, M. Yamaguchi, and T. Yanagida, Phys. Rev. Lett. {\bf
85}, 3572 (2000); Phys. Rev. D {\bf 63}, 103514 (2001).

\bibitem{murayama}
H. Murayama, H. Suzuki, T. Yanagida, and J. Yokoyama, Phys. Rev. D
{\bf 50}, 2356 (1994).

\bibitem{gaillard}
M.~K.~Gaillard, H.~Murayama, and K.~A.~Olive, Phys. Lett. B {\bf
355}, 71 (1995).

\bibitem{kkk2}
K. Enqvist, K.-W. Ng, and K. A. Olive, Nucl. Phys. B {\bf 303},
713 (1988).

\bibitem{byrnes}
C. T. Byrnes and D. Wands,  Phys. Rev. D {\bf 73}, 063509 (2006);
and references therein.

\bibitem{liddle}
A. Liddle and S. Leach, Phys. Rev. D {\bf 68}, 103503 (2003).

\bibitem{wmap}
H. Peiris {\it et al.}, Astrophys. J. Suppl. Ser. {\bf 148}, 213
(2003).

\bibitem{spergel}
D. N. Spergel {\it et al.}, Astrophys. J. Suppl. Ser. {\bf 170},
377 (2007).

\bibitem{hama}
K. Hamaguchi, M. Kawasaki, T. Moroi, and F.~Takahashi, Phys. Rev.
D {\bf 69}, 063504 (2004).

\bibitem{gluino}
A. Arvanitaki, C. Davis, P.~W. Graham, A. Pierce, and J.~G.
Wacker, Phys. Rev. D {\bf 72}, 075011 (2005).

\end{thebibliography}
\end{document}